# Exciton-photon interaction in a quantum dot embedded in a photonic microcavity

Majid Sodagar, Milad Khoshnegar, Amin Eftekharian, and Sina Khorasani

School of Electrical Engineering, Sharif University of Technology, P. O. Box 11365-9363, Tehran, Iran

**Abstract**
We present a detailed analysis of exciton-photon interaction in a microcavity made out of a photonic crystal slab. Here we have analyzed a disk-like quantum dot where an exciton is formed. Excitonic eigen-functions in addition to their eigen-energies are found through direct matrix diagonalization, while wave functions corresponding to unbound electron and hole are chosen as the basis set for this procedure. In order to evaluate these wave functions precisely, we have used $6 \times 6$ Luttinger Hamiltonian in the case of hole while ignoring bands adjacent to conduction band for electron states. After analyzing Excitonic states, a photonic crystal based microcavity with a relatively high quality factor mode has been proposed and its lattice constant has been adjusted to obtain the prescribed resonant frequency. We use finite-difference time-domain method in order to simulate our cavity with sufficient precision. Finally, we formulate the coupling constants for exciton-photon interaction both where intra-band and inter-band transitions occur. By evaluating a sample coupling constant, it has been shown that the system can be in strong coupling regime and Rabi oscillations can occur.

## 1. Introduction

Cavity quantum electrodynamics (CQED) has been the central topic of intense research since early 1990s, especially in optoelectronics and solid-state physics [1-7]. Recently, several experiments have been conducted in order to observe striking phenomena related to interactions occurring in such systems, and have addressed their possible applications in various regimes [8-14]. Generally speaking, in the weak coupling regime spontaneous emission can be enhanced or reduced compared with its vacuum level. However, in the strong coupling regime the key signature is Rabi splitting, which is observed in the emission spectrum in form of an anti-crossing between the quantum dot exciton and cavity-mode dispersion relations. In other words, Rabi oscillations occur in the decay dynamics, which take place before decoherence mechanisms.

Realization of single-photon sources as well as Q-bits, cryptography, quantum repeater, quantum computation and information are the main fields in which these phenomena seek application [11-17]. In most of these researches, photonic crystal (PhC) based cavities and waveguides are exploited owing to their favorable properties. There are several reports dealing with the design of PhC-based cavities with specific properties for special applications in CQED. These criteria include high quality factor and low mode volume, which are met through geometrical manipulations [18-22].

Exciton-photon interaction in a PhC waveguide and semiconductor quantum well (QW) as a two-dimensional system has been investigated theoretically in the literature [23,24]. Also the possibility of Rabi splitting predicted by theoretical models in strong coupling regime has been verified experimentally for semiconductor quantum dots in cavities [25-29]. A recent review paper by Reithmaier summarizes the progress made in the area of strongly-coupled nano-cavities, to which the reader is referred for more details [28].

We had made an initial attempt to investigate the quantum optical behavior of exciton-photon interaction in photonic crystals, where the interaction with Bloch modes were analyzed [30]. The authors had shown the formation of dressed states and Rabi oscillations, in strong-coupling regime. The present study, however, extends the latter research without making any approximations, taking the exact nature of excitonic wavefunctions as well as photonic confined modes into account. In this paper, we investigate the exciton-photon interaction in a disk-like quantum dot (QD) embedded in a 2D photonic crystal slab. The main motivation of this paper is to present a fundamental approach to design and understand the operation of such a cavity based on a basic theoretical method of the quantum optical and quantum mechanical phenomena involved. We show that it is possible to obtain an interaction within strong-coupling regime, which allows a wide range of applications as mentioned above. We discuss the detailed design of the quantum dot as well as the photonic crystal cavity. Finally, the photon-exciton coupling rate has been computed and dressed (entangled) exciton-photon states are found.

In section 2, we investigate excitonic states through diaonalization method and use unentangled electron and hole wave functions as the basis set. In section 3, a high quality factor PhC-based cavity has been designed, one of whose non-degenerate modes is selected, and its electric field spatial profile and its quality factor are obtained via FDTD method. Finally, we investigate the photon-exciton interaction in section 4 where the coupling coefficient for both intra-band and inter-band transitions are obtained.

## 2. Excitonic States

The electron-hole pair bound by Coulomb potential is called exciton, which is electrically neutral but can transport energy. Excitons can be formed by photon absorption at any critical point where the electron and hole group velocities are equal. Excitons can be classified into Frenkel or Wannier types according to the average electron-hole distance called exciton Bohr radius. Exciton's binding energy may be in the range of several to thousands of meV [31].

Frenkel excitons are localized near a single atom and hence have smaller Bohr radius while Wannier types are weaker in binding and have larger Bohr radius. In this case electron and hole reside in conduction and valence bands respectively. Henceforth, we will examine the Wannier exciton in this work. Although excitons can form complexes such as bi-exciton from two excitons, we merely confine ourselves to single exciton case. This situation can be practically met at cryogenic temperatures.

Excitons are formed not only in every insulating bulk crystals but also in low dimensional structures such as quantum wells, quantum wires, QDs and etc. where the crystal size is on the order of or less than the exciton Bohr radius. Here we are to achieve excitonic states in a disk-like QD which is 4nm thick and 150nm in radius. The dot and the barrier are supposed to be made from GaAs and $Ga_{0.74}Al_{0.36}As$ respectively so that electrons and holes experience a 300 meV and 150 meV potential walls, respectively. The single coupled electron-hole pair motion can be described by the following Hamiltonian

$$\widehat{H}_X = \widehat{H}_e + \widehat{H}_h + \widehat{V}_{e-h}, \qquad (1)$$

where $\widehat{H}_e$ and $\widehat{H}_h$ stand for the single particle Hamiltonians governing the electron and hole motions, respectively. Also, $\widehat{V}_{e-h}$ denotes the electron and hole Coulomb interaction potential which has the form

$$\widehat{V}_{e-h} = \frac{-e^2}{\varepsilon_r \varepsilon_0 |\mathbf{r}_e - \mathbf{r}_h|}. \qquad (2)$$

Here $e$ denotes the electronic charge while $\varepsilon_r$ and $\varepsilon_0$ stand for relative and vacuum permittivities, respectively. The relative permittivity corresponds to the material in which the electron-hole pair resides. In our case the material is GaAs and hence $\varepsilon_r = 13.2$. The surface polarization of the QD has been ignored in (1) [32]. The eigenvectors corresponding to single particle Hamiltonian $\widehat{H}_e$ and that of $\widehat{H}_h$ form a complete basis set for their own subspace. Hence every vector belonging to the exciton space on which $\widehat{H}_X$ operates, can be spanned by the tensor product of these eigenvectors. Particularly, the $p$th exciton Hamiltonian eigenvector $|\psi_p^X\rangle$ could be expanded in this space as

$$\left|\psi_p^X\right\rangle = \sum_{O,L} A_{O,L}^p \left|\psi_O^e\right\rangle \otimes \left|\psi_L^h\right\rangle. \qquad (3)$$

In the above expansion $\left|\psi_O^e\right\rangle$ and $\left|\psi_L^h\right\rangle$ stand for $O$th and $L$th eigenvectors correspond to electron and hole Hamiltonians, respectively. Notice that in a QD structure, $O$ and $L$ are collective indices, each standing for three quantum numbers. Also $A_{O,L}^p$ are the expansion coefficients. Equation (3) implies that an entangled electron-hole pair, i.e. exciton, exists only if at least two expansion coefficients be non zero. In view of the fact that our QD dimensions are big enough to encompass many atoms in all direction we are encouraged to assume envelope function approximation (EFA) in order to obtain single particle wave functions.

In the case of electron states, since the conduction band is not degenerate in non relativistic regime and other bands are far enough, the Bloch part of the wave function can be chosen as S-like orbital. So it is reasonable to ignore the adjacent bands, use only conduction band and simply solve the Schrödinger equation for a single envelope function.

It is straightforward to obtain the following projection of electron states on position bra owing to the simple QD structure

$$\left\langle r_e | \psi_O^e \right\rangle = A e^{in\varphi_e} J_n(\frac{\beta_{nv}\rho_e}{R_0}) Z_d^e(z_e) \chi^e(r_e). \qquad (4)$$

Here, $J_n$ is the Bessel's function of order n, $\beta_{nv}$ is the $v$th zero of $J_n$ and $R_0$ is the radius of our disk-like QD. Also $Z_d^e$ is the $d$th envelope function along the QD's height, i.e. $z$ direction, which is sinusoidal inside the dot and exponentially decaying outside. So $O$ stands for $n$, $v$ and $d$ quantum numbers, collectively. Also, $\chi$ stands for the S-like Bloch function and $A$ is normalization constant. The subscript $e$ is here used for the three position coordinates of electron, $\varphi$, $\rho$ and $z$ in cylindrical coordinate. Also $r_e$ stands for the aforementioned coordinate parameters collectively.

In contrast, the Bloch part corresponds to hole wave function mainly composed of $P_x$, $P_y$ and $P_z$-like orbital. The reason is that the valence band is not only twofold degenerate at Γ point due to heavy hole (HH) and light hole (LH) bands, but also the spin-orbit split-off (SO) band may be close to these bands in some semiconductor band structure and has a significant contribution in Bloch part of hole wave function.

In order to obtain hole wave functions accurately, one needs to take the effect of these neighboring bands into account. For the purpose of calculating envelope functions corresponding to each of these Bloch parts, Luttinger Hamiltonian could be exploited. It is possible to use Luttinger Hamiltonian in the cylindrical coordinate, because the radius of QD is an order of magnitude bigger that its thickness, we only employ it in the $z$ direction, i.e. along QD's height [33]. In $xy$-plane we still use single band approximation which results in similar outcomes to that of electrons. Also, we take the effects of only three adjacent bands, i.e. HH, LH and SO, into account and hence deal with the 6×6 Luttinger Hamiltonian. For more accuracy one can also take the conduction band into consideration and use 8×8 Luttinger Hamiltonian, but as far as the semi-conductor gap is large, the consequent improvement in modeling is negligible. This can readily be discerned from variational theory [34].
Irrespective of the method used to find the eigenvectors of Luttinger Hamiltonian, we can write the hole wave function as

$$\left\langle r_h | \psi_L^h \right\rangle = A e^{im\varphi_h} J_m(\frac{\beta_{m\omega}\rho_h}{R_0}) \sum_t Z_{b,t}^h(z_h) \chi_t^h(r_h). \qquad (5)$$

Here, the summation runs over the number of considered bands multiplied by two due to the spin of electrons and holes. Here $Z_{b,t}^h(z_h)$ is the $t$th component of the Luttinger Hamiltonian eigenvector corresponding to the $b$th bound envelope function along the QD's height. Other parameters are defined similar to that of (4), except that the index $e$ is replaced with $h$, indicating the position of hole instead of the electron. It is a common practice to diagonalize the $6\times 6$ Luttinger Hamiltonian matrix and achieve a $3\times 3$ block diagonal matrix through a unitary transformation [35-37].

$$\mathbf{H}_L^{6\times 6} = \begin{pmatrix} \mathbf{H}_L^+ & \mathbf{O} \\ \mathbf{O} & \mathbf{H}_L^- \end{pmatrix} \qquad (6)$$

In order to simplify the problem we also employ this procedure so that the reduced Hamiltonian reads

$$H_L^\pm = \begin{Bmatrix} P+Q+V & R\mp iS & \sqrt{2}R\pm i\sqrt{\frac{1}{2}}S \\ R\pm iS^\dagger & P-Q\mp iC+V & \sqrt{2}Q\mp i\sqrt{\frac{3}{2}}\Sigma \\ \sqrt{2}R\mp i\sqrt{\frac{1}{2}}S^\dagger & \sqrt{2}Q\pm i\sqrt{\frac{3}{2}}\Sigma^\dagger & P\pm iC+\Delta_{so}+V \end{Bmatrix}, \qquad (7)$$

Table 1. Luttinger Parameters for GaAs and AlAs

| Parameter | $\gamma_1$ | $\gamma_2$ | $\gamma_3$ | $\Delta_{so}$ (meV) |
|---|---|---|---|---|
| GaAs | 6.98 | 2.06 | 2.93 | 341 |
| AlAs | 3.76 | 0.82 | 1.42 | 280 |

in which $\Delta_{so}$ is the spin-orbit splitting at the $\Gamma$ point. Other parameters are given by

$$P = \left(\frac{\hbar^2}{2m_0}\right)\gamma_1\left(k_x^2 + k_y^2 + k_z^2\right), \tag{8a}$$

$$Q = \left(\frac{\hbar^2}{2m_0}\right)\gamma_2\left(k_x^2 + k_y^2 - 2k_z^2\right), \tag{8b}$$

$$R = -\sqrt{3}\left(\frac{\hbar^2}{2m_0}\right)\gamma_\varphi k_\parallel^2, \tag{8c}$$

$$S = 2\sqrt{3}\left(\frac{\hbar^2}{2m_0}\right)k_\parallel\left\{(\sigma - \delta)k_z + k_z\pi\right\}, \tag{8d}$$

$$\Sigma = 2\sqrt{3}\left(\frac{\hbar^2}{2m_0}\right)k_\parallel\left\{\left(\frac{1}{3}(\sigma-\delta)+\frac{2}{3}\pi\right)k_z \right.$$
$$\left. + k_z\left(\frac{2}{3}(\sigma-\delta)+\frac{1}{3}\pi\right)\right\}, \tag{8e}$$

$$C = 2\left(\frac{\hbar^2}{2m_0}\right)k_\parallel\left[k_z,(\sigma-\delta-\pi)\right], \tag{8f}$$

$$\gamma_\varphi = \sqrt{\overline{\gamma}^2 + \mu^2 - 2\overline{\gamma}\mu\cos(\varphi)}, \tag{8g}$$

$$\overline{\gamma} = \frac{\gamma_2 + \gamma_3}{2}, \quad \mu = \frac{\gamma_3 - \gamma_2}{2}, \tag{8h}$$

$$\sigma = \overline{\gamma} - \frac{1}{2}\delta, \quad \pi = \mu + \frac{3}{2}\delta, \tag{8i}$$

$$\delta = \frac{1}{9}(1 + \gamma_1 + \gamma_2 - 3\gamma_3), \tag{8j}$$

$$\varphi = \arctan\left(\frac{k_y}{k_x}\right). \tag{8k}$$

Also, [] stands for the commutation of two operators and $\gamma_i$ are the Luttinger parameters. Besides, $k_\parallel$ is the modulus of in plane wave vector, i.e. $k_\parallel = \sqrt{k_x^2 + k_y^2}$, and $m_0$ is the electron rest mass. Table 1 enumerates these parameters for two materials constructing our QD.

We have also employed linear interpolation for ternary composition $Ga_{0.74}Al_{0.36}As$ wherever applicable, and taken the influence of six strain components, i.e. $\varepsilon_{xx}$, $\varepsilon_{yy}$, $\varepsilon_{zz}$, $\varepsilon_{zx}$, $\varepsilon_{yz}$ and $\varepsilon_{xy}$, into account by adding the Pikus-Bir deformation potentials to the main Hamiltonian. It can be accomplished by adding the following potential to their corresponding counterpart [38], given by

$$P_\varepsilon = -a_v\left(\varepsilon_{xx} + \varepsilon_{yy} + \varepsilon_{zz}\right), \tag{9a}$$

$$O_\varepsilon = -\frac{b}{2}\left(\varepsilon_{xx} + \varepsilon_{yy} - 2\varepsilon_{zz}\right), \tag{9b}$$

$$R_\varepsilon = \frac{\sqrt{3}b}{2}\left(\varepsilon_{xx} - \varepsilon_{yy}\right) - id\varepsilon_{xy}, \tag{9c}$$

$$S_\varepsilon = -d\left(\varepsilon_{zx} - i\varepsilon_{yz}\right). \tag{9d}$$

Here, $a_v$, $b$ and $d$ are the Pikus-Bir deformation potentials describing the hydrostatic, uniaxial and shear strain, respectively. Owing to our rather planar QD structure we may take $\varepsilon_{yz} \approx \varepsilon_{zx}$, and ignore the effect of $S_\varepsilon$ or equivalently $d$. The corresponding parameters for GaAs and AlAs are listed in table 2 [39].

Since our QD can be considered as a layered structure in the z direction, we used transfer matrix method (TMM) in order to obtain the eigenstates of Hamiltonian (7) [40]. Note that the upper and lower blocks in Eq. (6), are related through complex conjugate operator, i.e. $\mathbf{H}^+ = (\mathbf{H}^-)^*$. Therefore, the eigenvectors of one, e.g. $\mathbf{H}^+$, can be achieved upon determination that of another through complex conjugate transformation. Henceforth we will focus on upper block eigenvalue problem, i.e. $\mathbf{H}^+\mathbf{F} = E\mathbf{F}$. It is straightforward to decompose $\mathbf{H}^+$ in three terms with respect to $k_z$

$$\mathbf{H}^+ = \mathbf{H}_2 k_z^2 + \mathbf{H}_1 k_z + \mathbf{H}_0 \tag{10}$$

If we assume constant potential profile in each layer, then all $k_z$ coefficients will be constant matrices. The eigenvalue problem results in three coupled second order differential equation in terms of $k_z$. It is possible to reduce the differential equation order by introducing $\mathbf{\Phi} = [\mathbf{F}, \mathbf{F}']^T$ at the cost of increasing the number of coupled equations to six. If so, the governing equation for $\mathbf{\Phi}$ reads

$$\mathbf{\Phi}' = \mathbf{\Lambda}\mathbf{\Phi} \tag{11a}$$

$$\mathbf{\Lambda} = \begin{pmatrix} \mathbf{O} & \mathbf{I} \\ \mathbf{H}_2^{-1}(\mathbf{H}_0 - E\mathbf{I}) & -i\mathbf{H}_2^{-1}\mathbf{H}_1 \end{pmatrix} \tag{11b}$$

Note that $\Lambda$ is not diagonal and hence the above equation cannot be solved directly. It can be decomposed as

$$\Lambda = PDP^{-1}$$

Here $D$ is a diagonal matrix composed of $\Lambda$ eigenvalues. Also $P$ is a square matrix composed of eigenvectors correspond to $\Lambda$. Notice that this decomposition is not unique. We choose $D$ so that its upper and lower half contains elements with positive and negative real part, i.e. forward and backward propagating waves, respectively. By introducing another change of variable, i.e. $Q = P^{-1}\Phi$, Eq. (11) recast in $Q' = DQ$ with the following solution

$$Q = e^D Q_0 \quad (12)$$

It is evident that envelope wave function and probability current continuity boundary conditions across an interface should be imposed on $\Phi$, rather than $Q$. Assuming that Bloch parts across interfaces remain unchanged, it is straightforward to write

$$B_L \Phi(z_L) = B_R \Phi(z_R) \quad (13)$$

Where the boundary condition matrix $B$ defined as

$$B = \begin{pmatrix} I & O \\ -iH_1 & -H_2 \end{pmatrix} \quad (14)$$

The total transfer matrix can be constructed by multiplying all transfer matrices correspond to each layer and applying appropriate boundary conditions across each interface. That is

$$(15) \quad T = \prod_{i=1}^{l-1} P_{i+1}^{-1} B_{i+1}^{-1} B_i P_i e^{D_i \Delta z_i}$$

Where $P_i$, $B_i$, $D_i$ and $\Delta z_i$ are the eigenvectors matrix, boundary condition matrix, eigenvalues matrix and length of $i$th layer. Also $l$ is the number of layers (here $l = 3$). As can be traced, the total transfer matrix will depend on system energy, i.e. $E$. In order to determine the permitted energies, wave function normalization should be considered.

Referring to Fig. 1, if $a_i$ and $b_i$ denote forward and backward wave vectors, respectively, at the interface between $i$th barrier and well, the following should hold for bound states, when $a_1$ is set to zero

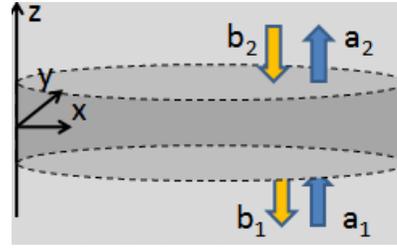

Figure 1. Forward and backward wave vectors depicted for the two barrier layers.

Table 2. Pikus-Bir Parameters for GaAs and AlAs

| $b$ | $a_v$ | |
|---|---|---|
| -2.0eV | -1.116eV | GaAs |
| -2.3eV | -2.47eV | AlAs |

$$(16) \quad \begin{aligned} a_2 &= T_{11}(E)a_1 + T_{12}(E)b_1 = T_{12}(E)b_1, \\ 0 &= T_{21}(E)a_1 + T_{22}(E)b_1 = T_{22}(E)b_1. \end{aligned}$$

So a particular $E$ is the system eigen-energy or equivalently the system has non-trivial bound state, if and only if $T_{22}$ has an eigenvalue equal to zero for that specific $E$. In order to find the eigenvalues, the energy $E$ can be swept while the determinant of $T_{22}$ is monitored. In figure 2 the outcome is depicted for a specific $k_\parallel$ near $\Gamma$ point in the reciprocal lattice.

The only problem involved with the above-mentioned approach is the appearance of spurious states. This is probable because such procedure is a perturbative, in which an incomplete basis set is used. These spurious states reveal themselves in a rather large eigenvalues of $\Lambda$ for some layer and hence may result in instability. This problem can be treated by eliminating such eigenvalues in $D$ matrix or clamping them appropriately.

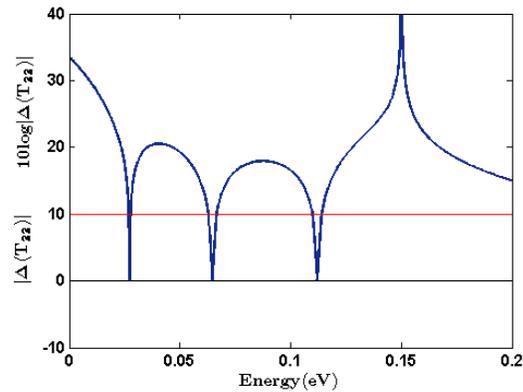

Figure 2. Transfer matrix determinant swept over energy. Zero-crossing points yield hole eigenenergies in the z direction.

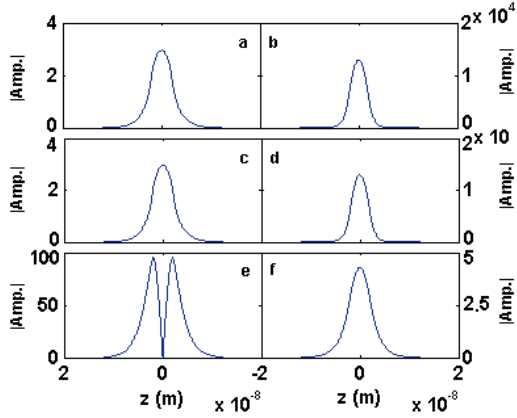

Figure 3. **a-f** envelope wave functions obtained for the first bound state correspond to $p_x\uparrow$, $p_x\downarrow$, $p_y\uparrow$, $p_y\downarrow$, $p_z\uparrow$ and $p_z\downarrow$ Bloch parts, respectively.

Figure 2 shows three zero crossing points in the swept range of energy, hence the system has at least three bound states. Note that for the sake of illustration two scales for vertical axis have been used. Also the corresponding eigenvector, i.e. $b_1$, can be evaluated for these three zero-crossing points, as shown in Fig. 3.

As the matrix elements of the Luttinger Hamiltonian in (7) are written in a symmetric basis other than pure π orbitals, i.e. $p_x\uparrow$, $p_x\downarrow$, $p_y\uparrow$, $p_y\downarrow$, $p_z\uparrow$ and $p_z\downarrow$, the resultant envelope wave functions $b_1$ are not symmetric with respect to the $z=0$ plane. Here, a unitary transformation is employed in order to symmetrize these envelope wave functions. In Fig. 3, the normalized six envelope wave functions $p_x\uparrow$, $p_x\downarrow$, $p_y\uparrow$, $p_y\downarrow$, $p_z\uparrow$ and $p_z\downarrow$, for the first bound state are depicted. It can be seen that the contribution of these envelope functions in total wave function is not identical. By investigating similar curves for other bound states it can be realized that the most significant envelope functions vary with respect to the bound state under consideration.

The matrix representation of exciton Hamiltonian operator, i.e. $\widehat{\mathbf{H}}_x^{r'r}$, is attained by projecting it on the space spanned by vectors obtained from tensor product of electron and hole Hamiltonian eigenvectors.

$$\widehat{\mathbf{H}}_x^{r'r} = \langle \psi^{r'} | \widehat{H}_x | \psi^r \rangle, \quad (17a)$$

$$|\psi^r\rangle = |\psi_O^e\rangle \otimes |\psi_L^h\rangle. \quad (17b)$$

The dimension of this matrix depends on the number of basis taken into account. Considering the exciton Hamiltonian in (1), one can recognize that in contrast with the first two terms, the third term, i.e. the Coulomb interaction term, results in integrations in a six dimensional space as the exciton Hamiltonian matrix elements are asked to evaluate.

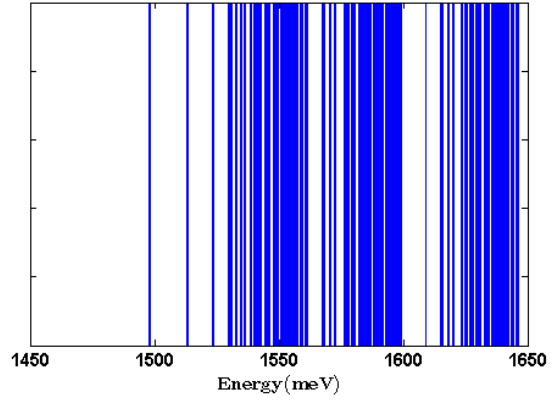

Figure 4. Spectrum of exciton energy obtained through matrix diagonalization. Each blue line denotes an exciton state.

Evaluation of such high dimensional integrals is computationally expensive, so we used IPM cluster computer equipped with LAM-MPI library [41,42]. Furthermore, we used VEGAS integration algorithm in order to increase the convergence rate [43,44]. The exciton Hamiltonian matrix representation is obtained by incorporating a basis set consisting of 243 ($3^5=243$) vectors. The number of bases depends on the chosen maximum of the six quantum numbers, i.e. $n$, $v$, $d$, $m$, $\omega$ and $b$. If we consider only bound states in $z$-direction, one can set $d_{max}=1$ and $b_{max}=3$ owing to the limited height of potential barriers in that direction for both electron and hole. Other quantum numbers correspond to the in-plane discretizations which have finer effect on energy distribution comparing to $b$ and $d$. That is why the QD radius is an order of magnitude bigger that its thickness. Also, $n_{max}$, $m_{max}$, $\omega_{max}$ and $v_{max}$ have been set to 3. This choice can contain more bases at the cost of more computation burden, but yield negligible refinement for our purpose. Since the exciton Hamiltonian is Hermitian, we needed to evaluate only 29646 matrix elements. In Fig. 4, the calculated spectrum of exciton Hamiltonian eigen-energies is shown from direct diagonalization of its matrix representation. Also the $A_{O,L}^p$ coefficients in (3) are obtained through this analysis by evaluating the corres-ponding eigenvectors.

## 3. Photonic Crystal Cavity

In order to have a strong coupling between exciton and photon, one needs to trap the photon in a rather high quality factor cavity, in which photon cannot escape easily. PhC-based cavities seem to be the best solution thanks to their unique abilities to manipulate light. There are several papers published in designing high quality factor PhC based cavities [19,21,45-48]. Although the 3D-PhC cavities are more effective in light trapping, 2D-PhC slab cavities are easier to fabricate, and we therefore will deal with such 2D-slab configurations.

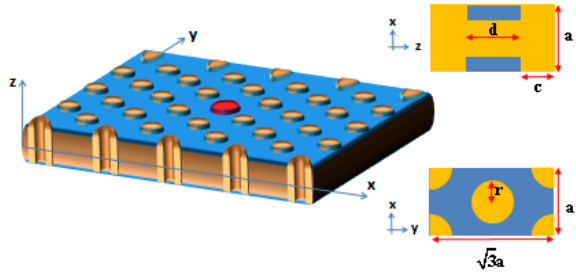

Figure 5. PhC Slab cavity structure formed by introducing a point defect.

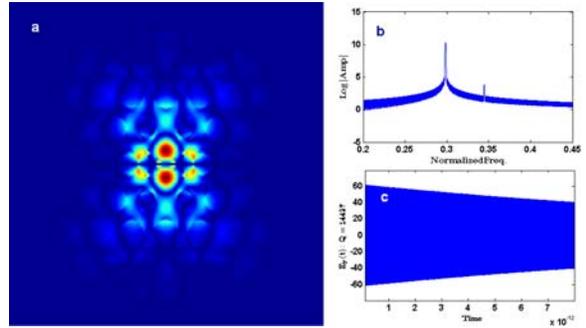

Figure 6. a) Mode spatial profile ($E_y$ component) accompanied with its time domain (c) responses. b) Cavity frequency response excited with a broadband signal. Symmetric boundary conditions along $x=0$ and $y=0$ planes has been employed.

We are interested in a high quality factor TE-like mode with small mode volume. The microcavity is formed by introducing a point defect in a perforated air hole triangular PhC slab depicted in Fig. 5.

In such structures, total internal reflection and Bragg reflection are the two mechanisms responsible for light localization in the out-of-plane and in-plane directions, respectively. In this structure, delocalization and cancelation mechanisms can be applied to decrease the vertical losses [49-51]. Since the structure is symmetric with respect to $z=0$ plane, TE-Like and TM-Like modes are available. Also the TE gap occurs between two lowest bands. Choosing slab thickness and air hole radius in the range of $0.5a < d < 1.2a$ and $0.35a < r < 0.4a$ respectively, yields a relatively large gap over mid-gap ratio. In our design, we set $d = 0.71a$ and $r = 0.37a$ which results in opening a gap between the normalized frequencies 0.2574 and 0.3911.

The defect modes frequencies are obtained through finite-difference time-domain (FDTD) analysis by exciting the cavity with a broadband electric-dipole having bandwidth equal to PhC band gap. Yee's algorithm is employed to implement FDTD [52]. We will focus on a symmetric mode both with respect to $x=0$ and $y=0$ plane which has the lowest frequency. Furthermore, the radius of nearest air holes is set to $0.3a$ in order to achieve a relatively high quality factor as well as small mode volume [21]. The increase in quality factor by changing the radius of the nearest holes is owing to the redistribution of the spatial Fourier components, i.e. in k-space, of the cavity mode electric field. In fact this trick results in pushing the Fourier components out of the light cone and hence reduces the vertical radiation losses.

The corresponding Q-factor is estimated by exciting the cavity with a narrowband electric-dipole centered at mode resonant frequency and monitoring the electric field decay at a low symmetric point after the input is switched off. The frequency and time domain responses as well as mode spatial profile are depicted in Fig. 6.

For obtaining reasonable results, Split-PML absorbing boundary condition is applied at the top and bottom layers each shifted about $0.5a$ into air, i.e. $c$ is set to $0.5a$ [53]. Also four periods in each direction is employed and seemed sufficient. The grid is set to 20 cells per lattice constant, and subpixel smoothing is employed for better accuracy [54].

The Q-factor and normalized resonant frequency are evaluated as about 14,000 and 0.298, respectively, for the probed mode. Note that the experimental cavity Q-factor is expected to be slightly lower than that of simulated due to other loss mechanisms such as material absorption and scattering from fabrication imperfections. We will choose PhC lattice constant, $a$, such that the exciton and photon frequencies be in resonance. Also, the PhC mode Q-factor can be used to approximate the density of states when system time evolution is under consideration.

## 4. Exciton-Photon Interaction

In this section we investigate the interaction between exciton and photon residing in the aforementioned QD and PhC cavity, respectively. We will limit ourselves by supposing only one photon resides in the cavity. The governing Hamiltonian for the system consists of exciton and photon can be simplified by using rotating wave approximation (RWA), i.e. supposing energy conservation is hold during interaction. In the case of exciton Hamiltonian, we will exploit second quantized form and field operators in order to work in a consistent mathematical framework. The total Hamiltonian reads

$$\widehat{H} = \widehat{H}_X + \widehat{H}_{EM} + \widehat{H}_i, \tag{18a}$$

$$\widehat{H}_X = \sum_p \mathcal{E}_p u_p^\dagger u_p, \tag{18b}$$

$$\widehat{H}_{EM} = \sum_{\mathbf{k}} \hbar \omega_{\mathbf{k}} c_{\mathbf{k}}^\dagger c_{\mathbf{k}}. \tag{18c}$$

Here, $\mathcal{E}_p$ and $\omega_{\mathbf{k}}$ denote the exciton energy at level $p$ and photon temporal frequency in $k$th mode, respectively. Also $c_{\mathbf{k}}^\dagger$ and $c_{\mathbf{k}}$ are bosonic creator and annihilator of photon at $k$th mode satisfying the commutation relationship $\left[c_{\mathbf{k}}, c_{\mathbf{k}}^\dagger\right] = 1$, respectively, Similarly, $u_p^\dagger$ and $u_p$ stand for fermionic creator and annihilator of exciton at level $p$ satisfying the anticommutator relationship $\{u_p, u_p^\dagger\} = 1$. Based on (3), it is straightforward to observe that

$$u_p^\dagger = \sum_{O,L} A_{O,L}^p a_O^\dagger b_L^\dagger, \tag{19a}$$

$$u_p = \sum_{O,L} A_{O,L}^p a_O b_L, \tag{19b}$$

$$\left|\psi_p^X\right\rangle = u_p^\dagger \left|0_X\right\rangle, \tag{19c}$$

where $a_O^\dagger$ and $b_L^\dagger$ are creators of electron and hole at $O$th and $L$th levels, respectively. Here $\left|0_X\right\rangle$ denotes the excitonic vacuum state in which there is no electron-hole pair.

Minimal coupling and direct coupling schemes are the two equivalent approaches for modeling of the interaction Hamiltonian [55]. Hereafter, we will assume minimal coupling scheme. The interaction of photon and exciton may result in two physically distinct processes which should be dealt with separately. The process in which exciton is created or annihilated, i.e. when excitonic vacuum state $\left|0_X\right\rangle$ is involved, and process in which the exciton level changes. The former consists of an electronic transition from valence to conduction band, i.e. intra-band transition, while the latter consists of only inter-band transitions of electrons and holes.

Regarding intra-band transition the interaction Hamiltonian will be linear in terms of excitonic and photonic field operators. In this case the interaction Hamiltonian reads

$$\widehat{H}_i = \sum_{p,\mathbf{k}} g_{p,\mathbf{k}} u_p^\dagger c_{\mathbf{k}} + \text{H.c.}, \tag{20}$$

where $g_{p,k}$ denotes the coupling coefficient between $k$th electromagnetic and $p$th excitonic states, and H.c. represents the Hermitian conjugate.

By exploiting minimal coupling scheme and neglecting second order terms which is justified in low intensity processes, the first quantization form of $\widehat{H}_i$ can simplified as

$$\widehat{H}_i = \frac{e}{2m_e} \sum_{j=1}^N \left[\mathbf{A}(\mathbf{r}_j) \cdot \mathbf{p}_j + \mathbf{p}_j \cdot \mathbf{A}(\mathbf{r}_j)\right]. \tag{21}$$

The summation is over all unit cells in the QD semiconductor crystal, supposing there is only one electron per unit cell. Here, $\mathbf{p}_j$ is the $j$th electron momentum operator while $m_e$ and $e$ are the electron rest mass and charge, respectively. Also, $\mathbf{A}$ represents the vector potential operator given by

$$\mathbf{A}(\mathbf{r}) = \sum_{\mathbf{k}} \sqrt{\frac{\hbar}{2\varepsilon_0 \omega_{\mathbf{k}} V}} \left(\mathbf{E}_{\mathbf{k}}(\mathbf{r}) c_{\mathbf{k}} + \mathbf{E}_{\mathbf{k}}^*(\mathbf{r}) c_{\mathbf{k}}^\dagger\right). \tag{22}$$

Here, $\mathbf{E}_{\mathbf{k}}(\mathbf{r})$ is $k$th mode spatial profile. By employing dipole approximation as well as RWA one can obtain the coupling coefficient as

$$g_{\mathbf{k},p} = -ie\omega_X^p \sqrt{\frac{\hbar}{2\varepsilon_0 \omega_{\mathbf{k}} V}} \mathbf{E}(\mathbf{r}_0) \cdot \tilde{\mathbf{g}}, \tag{23a}$$

$$\tilde{\mathbf{g}} \approx \sum_{L,O,t} A_{L,O}^p \left\langle \Psi_O^e | \Psi_{L,t}^h \right\rangle \left\langle \chi^e | \mathbf{r} | \chi_t^h \right\rangle, \tag{23b}$$

where $\Psi$ stands for envelope part of wave function. The detailed derivation of the above relationship is given in the Appendix.

As it was mentioned in section 2, $\chi$ is a linear combination of π-like orbitals for hole wave function, and an S-like orbital for electron wave function. It is a common practice to employ experimental data to evaluate the above electric dipole moment matrix elements, because the involving Bloch parts are not exactly known for various semiconductor crystals. In the case of GaAs the matrix element $\langle p_d | d | s \rangle$ is on the order of 30 Debye in which $d$ stands for either $x$, $y$ or $z$ [56,57]. Other matrix elements are zero owing to symmetry.

In contrast to intra-band transitions, the interaction Hamiltonian corresponds to the inter-band transition is bilinear in terms of excitonic field operators. In this case the second quantization form of the interaction Hamiltonian can be written as

$$\widehat{H}_i = \sum_{p',p,\mathbf{k}} g_{p,p',\mathbf{k}} u_p^\dagger u_p c_{\mathbf{k}} + \text{H.c.} . \quad (24)$$

In this case, the coupling coefficients are defined through the following matrix element

$$g_{p,p',\mathbf{k}} = \langle \psi_X^p, 0_{\mathbf{k}} | \widehat{H}_i | \psi_X^{p'}, 1_{\mathbf{k}} \rangle . \quad (25)$$

Again, exploiting minimal coupling scheme and neglecting second order terms, yields an expression similar to (21) as

$$\widehat{H}_i = \frac{1}{2} \sum_{w \in \{e,h\}} \frac{e_w}{m_w} \left[ \widehat{\mathbf{A}}(\mathbf{r}_w) \cdot \widehat{\mathbf{p}}_w + \widehat{\mathbf{p}}_w \cdot \widehat{\mathbf{A}}(\mathbf{r}_w) \right], \quad (26)$$

where the summation is over electron and hole from which an exciton is constituted. After substituting (26) and (22) into (25), the latter can be reduced to

$$g_{p,p',\mathbf{k}} = g_{p,p',\mathbf{k}}^e + g_{p,p',\mathbf{k}}^h \quad (27a)$$

$$g_{p,p',\mathbf{k}}^e = \frac{ie\hbar}{m_e} \sqrt{\frac{\hbar}{2\varepsilon_0 \omega_{\mathbf{k}} V}} \mathbf{E}_{\mathbf{k}}(\mathbf{r}_{e0}) \cdot \widetilde{\mathbf{g}}^e \quad (27b)$$

$$\widetilde{\mathbf{g}}^e = \sum_{O,L,O'} A_{O,L}^y A_{O',L}^z \int \psi_e^{O*}(\mathbf{r}_e) \nabla_e \psi_e^O(\mathbf{r}_e) d^3 \mathbf{r}_e \quad (27c)$$

Again, dipole approximation is employed. The counterpart equations for (27b) and (27c), i.e. for hole couplings, can be achieved by replacing $e$, $L$ and $O$ indices by $h$, $O$ and $L$, respectively. It is possible to further simplify the integrand by employing EFA and rewrite electron and hole functions in terms of corresponding envelope and Bloch parts.

In order to investigate the interaction phenomenon, we will simplify the total Hamiltonian (18) by considering only one excitonic state interacting with only one photonic mode. Furthermore, we assume that intra-band transition occurs during the interaction, and hence use the linear representation in (20). This gives

$$\widehat{H} = \mathcal{E}_p u_p^\dagger u_p + \hbar \omega_{\mathbf{k}} c_{\mathbf{k}}^\dagger c_{\mathbf{k}} + g_{p,\mathbf{k}} u_p^\dagger c_{\mathbf{k}} + \text{H.c.} . \quad (28)$$

Here the cavity mode has a finite Q-factor and hence a complex frequency. Henceforth $\gamma_c$ denotes the full width at half maximum (FWHM) of the cavity mode.

Phenomenologically, the nonradiative exciton decay due to electron-phonon interactions can be included with a nonradiative linewidth, denoted by $\gamma_X$, which allows the QD to be described via a complex frequency. This broadening is not radiatively limited, but is due to dephasing mechanisms, such as Coulomb interaction with free carriers [58]. Exciton nonradiative linewidth is on the order of few tens of µeV [59].

Considering these effects, one can find the Hamiltonian eigen-energies which reads [60,61]

$$\omega^\pm = \omega_0 - i\gamma_o \pm \sqrt{g^2 - \left(\frac{\gamma_c - \gamma_X}{4} - i\frac{\Delta}{2}\right)^2}, \quad (29a)$$

$$\omega_0 = \frac{1}{2}(\omega_X + \omega_c), \quad (29b)$$

$$\gamma_o = \frac{1}{4}(\gamma_c + \gamma_X), \quad (29c)$$

in which $\Delta$ stands for detuning between exciton and photon states. When the coupling coefficient is large enough so that there are two eigenfrequencies with distinct real parts correspond to two non-degenerate entangled states, the system operates in a strong coupling regime. We set the PhC lattice constant to 244nm so that the cavity mode under consideration be in resonance with the third exciton state, i.e. $\Delta = 0$. The corresponding energy is 1513.3 meV. Also the cavity mode linewidth $\gamma_c$ is calculated as 110µeV equivalent to its quality factor. The QD is assumed to be located at the peak of photonic mode's electric field profile in order to maximize the coupling coefficient. In order to have Rabi splitting, or equivalently the system to be in strong coupling regime, it should has eigen-frequencies with distinct real values. Equation (29) indicates that when cavity mode and exciton state are at resonance, the sufficient condition for the system to be in strong coupling regime is $g^2 > \frac{1}{16}(\gamma_c - \gamma_X)^2$. By using equation (23) and the results obtained in section 3, coupling coefficient is evaluated about 159 GHz. This predicts that Rabi oscillation will occur at $2\sqrt{g^2 - \frac{1}{16}(\gamma_c - \gamma_X)^2}$ Hz.

## 5. Conclusion

In this paper, we studied the bound electron-hole pairs, i.e. excitons, quantum mechanically and used Luttinger Hamiltonian in order to achieve their eigenstates in a disk-like QD precisely. We exploited these states to construct a basis set for the Hilbert space in which exciton lives and use matrix diagonalization method to approximate exciton eigenstates. Afterwards a very high quality factor cavity has been designed by introducing a point defect in a triangular lattice photonic crystal slab. The coupling coefficient between photon and exciton states was examined and reformulated in order to inspect the interaction phenomenon. This phenomenon, in the field of quantum electro-dynamics, quantum information and quantum computing has a variety of applications and plays a key role both in weak and strong coupling regimes. Finally, the numerical value of coupling coefficient corresponding to a single exciton state and a high quality photon mode was evaluated and shown that the system is capable of operating in the strong coupling regime.

**Acknowledgement**

The authors gratefully acknowledge the support of Nanotechnology Supercomputing Center at the Institute for Physics and Mathematics (IPM, P. O. Box 19395-5531, Tehran, Iran) for providing the hardware facility.

This work has been supported in part by Iranian National Science Foundation (INSF).

**Appendix**

In this appendix we will derive coupling coefficient expressions for intra-band transitions. The derivation of coupling coefficient corresponding to inter-band transition is very similar. In case of intra-band transition the interaction Hamiltonian is

$$\widehat{H}_i = \frac{e}{2m_e}\sum_{j=1}^{N}\left[\mathbf{A}(\mathbf{r}_j)\cdot\mathbf{p}_j + \mathbf{p}_j\cdot\mathbf{A}(\mathbf{r}_j)\right]. \quad (A.1)$$

We substitute the momentum operator using $i\hbar\mathbf{p}_j = m_e[\mathbf{r}_j, H_e]$ and further simplify $H_i$ by supposing $[\mathbf{A}(\mathbf{r}_j), H_e] \approx 0$, which is justified by assuming dipole approximation. Hence, we get

$$\mathcal{H}_{int} = \frac{e}{i\hbar}\sum_{i=1}^{N}[\mathbf{r}_i, H_e]\cdot\mathbf{A}(\mathbf{r}_i) \quad (A.2)$$

in which $H_e$ corresponds to interacting electrons in semiconductor given by

$$H_e = \sum_{i=1}^{N}\frac{\mathbf{p}_i^2}{2m} + \frac{1}{2}\sum_{i\neq j}\frac{e^2}{|\mathbf{r}_i - \mathbf{r}_j|} - \sum_{i,l}\frac{Ze^2}{|\mathbf{r}_i - \mathbf{R}_l|}, \quad (A.3)$$

where $i$ and $l$ refer to the index of the electron and nuclei, respectively. Also, the quantized vector potential for photonic modes is

$$\mathbf{A}(\mathbf{r}) = \sum_{\mathbf{k}}\sqrt{\frac{\hbar}{2\varepsilon_0 \omega_{\mathbf{k}} V}}\left(\mathbf{E}_{\mathbf{k}}(\mathbf{r})c_{\mathbf{k}} + \mathbf{E}_{\mathbf{k}}^*(\mathbf{r})c_{\mathbf{k}}^\dagger\right). \quad (A.4)$$

By exploiting two-band semiconductor model, one can write the many-electron ground state with the aid of the Slater determinant

$$\varPsi_0 = \mathcal{A}\left\{\psi_{vk_1}, \psi_{vk_2}, \ldots, \psi_{vk_i}, \ldots, \psi_{vk_N}\right\}. \quad (A.5)$$

Here, $\mathcal{A}$ is the antisymmetrizing operator, $N$ is the number of unit cells, and $\psi_{vk_i}$ is the one-electron wave function corresponds to $i$th electron in the semiconductor valence band. Also the exciton state at level $p$ can be written as

$$|\varPsi_p\rangle = \mathcal{A}\left\{\psi_{vk_1}, \psi_{vk_2}, \ldots, \psi_X^p, \ldots, \psi_{vk_N}\right\}. \quad (A.6)$$

Note that $|\varPsi_p\rangle$ is the eigenket of $H_e$ with eigen energy $\mathcal{E}_p$. The coupling constant is defined through the following matrix element

$$g_{\mathbf{k},p} = \langle\varPsi_p, \mathbf{0}^{ph}|\mathcal{H}_{int}|\varPsi_0, \mathbf{1}^{ph}\rangle. \quad (A.7)$$

It is straightforward to reduce (A.7) after substituting equations (A.2), (A.4) and (3) by using bosonic creator and annihilator properties as

$$g_{\mathbf{k},p} = \frac{e}{i\hbar}\sqrt{\frac{\hbar}{2\varepsilon_0 \omega_{\mathbf{k}} V}} \mathcal{E}_p \hat{g}, \qquad (A.8a)$$

$$\begin{aligned}\hat{g} &= \langle \mathscr{P}_p | \sum_{i=1}^N \mathbf{E}(\mathbf{r}_i)\cdot \mathbf{r}_i | \mathscr{P}_0 \rangle \\ &= \sum_{L,O} A_{L,O}^p \langle \psi_e^L | \mathbf{E}(\mathbf{r})\cdot \mathbf{r} | \psi_h^O \rangle\end{aligned} \qquad (A.8b)$$

since electric filed profile has a large spatial wavelength compare to that of electronic parts, one can ignore electric filed variations and exclude it form integration. Upon using (EFA) in case of electronic wave functions, equation (A.8b) is further simplified as

$$g_{\mathbf{k},p} = -ie\omega_X^p \sqrt{\frac{\hbar}{2\varepsilon_0 \omega_k V}} \mathbf{E}(\mathbf{r}_0)\cdot \tilde{\mathbf{g}} \qquad (A.9a)$$

$$\tilde{\mathbf{g}} \approx \sum_{L,O,t} A_{L,O}^p \langle \Psi_L^e | \Psi_{O,t}^h \rangle \langle \chi^e | \mathbf{r} | \chi_t^h \rangle \qquad (A.9b)$$

## References

[1] Benisty H, Gérard J M, Houdré R, Rarity J and Weisbuch C 1999 *Confined Photon Systems: Fundamentals and Applications* (Series: *Lecture Notes in Physics*, vol **531**) (Berlin: Springer)
[2] Weisbuch C and Rarity J 1996 *Microcavities and Photonic Bandgaps: Physics and Applications* (NATO ASI Series *E* vol **324**) (Kluwer: Dordrecht)
[3] Burstein E and Weisbuch C 1995 *Confined Electrons and Photons: New Physics and Applications* (NATO ASI Series *B* vol **340**) (Plenum: New York)
[4] Ducloy M and Bloch D 1996 *Quantum Optics of Confined Systems* (NATO ASI Series *E* vol **314**) (Kluwer: Dordrecht)
[5] Yamamoto Y and Slusher R E 1993 Optical processes in microcavities *Physics Today* **46**, 66–73
[6] Gerard J M and Gayral B 2001 InAs quantum dots: artificial atoms for solid-state cavity-quantum electrodynamics *Physica* E **9** 131–139
[7] Vahala K J 2003 Optical microcavities *Nature* **424** 839–846
[8] Julsgaard B, Johansen J, Stobbe S, Stolberg-Rohr T, Sünner T, Kamp M, Forchel A and Lodahl P 2008 Decay dynamics of quantum dots influenced by the local density of optical states of two-dimensional photonic crystal membranes *Appl. Phys. Lett.* **93** 094102
[9] Hennessy K, Badolato A, Winger M, Gerace D, Atature M, Gulde S, Falt S, Hu E L and Imamoglu A 2007 Quantum nature of a strongly coupled single quantum dot–cavity system *Nature* **445** 896-899
[10] Kuroda T et al, 2008 Acceleration and suppression of photoemission of GaAs quantum dots embedded in photonic crystal microcavities *Appl. Phys. Lett.* **93** 111103
[11] Kaniber M et al 2008 Tunable single quantum dot nanocavities for cavity QED experiments *J. Phys.: Condens. Matter* **20** 454209
[12] Inoue J, Ochiai T and Sakoda K 2008 Spontaneous emission properties of a quantum dot in an ultrahigh-Q cavity: Crossover from weak-to-strong-coupling states and robust quantum interference *Phys. Rev.* A **77** 015806
[13] Kaniber M and Laucht A 2008 Investigation of the nonresonant dot-cavity coupling in two-dimensional photonic crystal nanocavities *Phys. Rev.* B **77** 161303
[14] Aoki K, Guimard D, Nishioka M, Nomura M, Iwamoto S and Arakawa Y 2008 Coupling of quantum-dot light emission with a three-dimensional photonic crystal nanocavity *Nature* **2** 688-692
[15] Lounis B and Orrit M 2005 *Rep. Prog. Phys.* **68** 1129
[16] Knill E, Laflamme R and Milburn G J 2001 *Nature* **409** 46-52
[17] Giovannetti V, Lloyd S and Maccone L 2008 Quantum Random Access Memory *Phys. Rev. Lett.* **100** 160501
[18] Frei R, Johnson H T and Choquette D 2008 Optimization of a single defect photonic crystal laser cavity *J. Appl. Phys.* **103** 033102
[19] Akahane Y, Asano T, Song B Sh and Noda S 2003 High-Q photonic nanocavity in a two-dimensional photonic crystal *Nature* **425** 944-947
[20] Vuckovic J and Yamamoto Y 2003 Photonic crystal microcavities for cavity quantum electrodynamics with a single quantum dot *Appl. Phys. Lett.* **82** 2374
[21] Ryua H Y and Notomi M 2003 High quality factor and small mode volume hexapole modes in photonic crystal slab nanocavities *Appl. Phys. Lett.* **83** 4294-4296
[22] Kuramochi E, Taniyama H, Tanabe T, Shinya A and Notomi M 2008 Ultrahigh-Q two-dimensional photonic crystal slab nanocavities in very thin barriers *Appl. Phys. Lett.* **93** 111112
[23] Yang Sh and John S 2007 Exciton dressing and capture by a photonic band edge *Phys. Rev.* B **75** 235332
[24] Gerace D and Andreani L C 2007 Quantum theory of exciton-photon coupling in photonic crystal slabs with embedded quantum wells *Phys. Rev.* B **75**, 235325
[25] Reithmaier J P Sek G, Loffler A, Hofmann C, Kuhn S, Reitzenstein S, Keldysh L V, Kulakovskii V D, Reinecke T L and Forchel A 2004 Strong coupling in a single quantum dot–semiconductor microcavity system *Nature* **432** 197-200
[26] Yoshie T, Scherer A, Hendrickson J, Khitrova G, Gibbs H M, Rupper G, Ell C, Shchekin O B and D. G. Deppe 2004 Vacuum Rabi splitting with a single quantum dot in a photonic crystal nanocavity *Nature* **432** 200-203
[27] Aoki K, Guimard D, Nishioka M, Nomura M, Iwamoto S AND Arakawa Y 2008 coupling of quantum-dot light emission with a three-dimensional photonic crystal nanocavity *Nature* **2** 688-692
[28] Reithmaier J P 2008 Strong exciton–photon coupling in semiconductor quantum dot systems *Semicond. Sci. Technol.* **23** 123001
[29] Ota Y, Shirane M, Nomura M, Kumagai N, Ishida S, Iwamoto S, Yorozu S, and Arakawa Y 2009 Vacuum Rabi splitting with a single quantum dot embedded in a H1 photonic crystal nanocavity *Appl. Phys. Lett.* **94** 033102
[30] Sodagar M, Khorasani S, Atabaki A, and Adibi A 2008 Quantum optics of a quantum dot embedded in a photonic crystal cavity *Proc. SPIE* **6901** 69010E
[31] Kittel C 2005 *Introduction to Solid State Physics* (John Wiley & Sons) 435-436
[32] Banyai L, Hu Y Z, Lindberg M and Koch S. W 1988 Third-order optical nonlinearities in semiconductor microstructures *Phys. Rev.* B **38** 8142-8153

[33] García-Cristóbal A, Fomin V M and Devreese J T 1998 Electronic structure of self-assembled quantum dots in high magnetic fields *Physica B* **256** 190-193
[34] Sakurai J J 2005 Modern Quantum Mechanics (Addison Wesley) 285-293
[35] Ahna D and Yoon S J 1995 Theory of optical gain in strained-layer quantum wells within the 6x6 Luttinger-Kohn model *J. Appl. Phys.* **78** 4
[36] Lopez-Villanueva A, Melchor I and Carceller J E 1999 Hole confinement and energy subbands in a silicon inversion layer using the effective mass theory *J. Appl. Phys.* **86** 1
[37] Gamiz F, Palma A and Cartujo P 2000 Influence of technological parameters on the behavior of the hole effective mass in SiGe structures *J. Appl. Phys.* **88** 4
[38] Chao C Y P and Chuang S L 1992 Spin-orbit-coupling effects on the valence-band structure of strained semiconductor quantum wells *Phys. Rev.* B **46** 4110
[39] Harrison P 2005 *Quantum Wells, Wires and Dots* (2nd ed: Wiley)
[40] Chen B, Lazzouni M and Ram-Mohan L R 1992 Diagonal representation for the transfer-matrix method for obtaining electronic energy levels in layered semiconductor heterostructures *Phys. Rev.* B **45** 1204
[41] http://nano.ipm.ac.ir/
[42] http://www.lam-mpi.org/
[43] Lepage G P 1978 A New algorithm for adaptive multidimensional integration *J. Computational Phys.* **27** 192–203
[44] Lepage G P 1980 VEGAS: An adaptive multidimensional integration program *Publication CLNS* **80** 447
[45] Song B S, Noda S, Asano T and Akahane Y 2005 Ultra-high-Q photonic double heterostructure nanocavity *Nat. Mater.* **4** 3 207–210
[46] Chen L and Towe E 2006 Design of high-Q microcavities for proposed two-dimensional electrically pumped photonic crystal lasers *IEEE J. Sel. Topics Quantum Elect.* **12** 1
[47] Frei W R, Johnsona H T and Choquette K D 2008 Optimization of a single defect photonic crystal laser cavity *J. Appl. Phys.* **103** 033102
[48] Fan Sh, Villeneuve P R, Joannopoulos J D and Haus H A 1998 Channel drop tunneling through localized states", *Phys. Rev. Lett.* **80** 960
[49] Fan Sh, Villeneuve P R, Joannopoulos J D and Haus H A 2001 Loss-induced on/off switching in a channel add/drop filter *Phys. Rev.* B **64** 245302
[50] Benisty H, Labilloy D, Weisbuch C, Smith C J M, Krauss T F, Cassagne D, Béraud A and Jouanin C 2000 Radiation losses of waveguide-based two-dimensional photonic crystals: positive role of the substrate *Appl. Phys. Lett.* **76** 532
[51] Johnson S G, Fan Sh, Mekis A and Joannopoulos J D 2001 Multipole-cancellation mechanism for high-Q cavities in the absence of a complete photonic band gap *Appl. Phys. Lett.* **78**, 3388
[52] Yee K S 1966 Numerical solution of initial boundary value problems involving Maxwell's equations in isotropic media *IEEE Trans. Antennas Propagat.* **14** 302
[53] Berenger J P 2007 Perfectly Matched Layer (PML) for Computational Electromagnetics (Morgan & Claypool)
[54] Farjadpour A, Roundy D, Rodriguez A, Ibanescu M, Bermel P, Joannopoulos J D and Johnson S G 2006 Improving accuracy by subpixel smoothing in the finite-difference time domain *Opt. Lett.* **31** 20
[55] Rewski K and Boyd R W 2004 Equivalence of interaction Hamiltonians in the electric dipole approximation *J. Mod. Opt.* **5** 8 1137–1147
[56] Silverman K L, Mirin R P, Cundiff S T and Norman A G 2003 Direct measurement of polarization resolved transition dipole moment in InGaAs/GaAs quantum dots *Appl. Phys. Lett.* **82** 25
[57] Wu D C, Kao J K, Mao M H, Chang F Y and Lin H H 2007 Determination of interband transition dipole moment of InAs/InGaAs quantum dots from modal absorption spectra *Proc. CLEO* **6** 1-2
[58] Kammerer C, Voisin C, Cassabois G, Delalande C, Roussignol Ph, Klopf F, Reithmaier J P, Forchel A and Gérard J M 2002 Line narrowing in single semiconductor quantum dots: Toward the control of environment effects *Phys. Rev.* B **66**, 041306
[59] Peter E, Senellart P, Martrou D, Lemaîtrе A, Hours J, Gerard J M, and Bloch J 2005 Exciton-photon strong-coupling regime for a single quantum dot embedded in a microcavity *Phys. Rev. Lett.* **95** 067401
[60] Andreani L, Panzarini G and Gerard J M 1999 Strong-coupling regime for quantum boxes in pillar microcavities: Theory *Phys. Rev.* B **60** 13276–13279
[61] Rudin S and Reinecke T L 1999 Oscillator model for vacuum Rabi splitting in microcavities. *Phys. Rev.* B **59** 10227–1023